\documentclass[twocolumn]{aastex631}

\usepackage{amsmath}
\usepackage{amssymb}
\usepackage{bm}

\begin{document}

\title{Standing solitary waves as transitions to spiral structures in \\ gravitationally unstable accretion disks}

\author{Hongping Deng}
\email{hpdeng353@gmail.com}
 \affiliation{Department of Applied Mathematics and Theoretical Physics, University of Cambridge, Centre for Mathematical Sciences, Wilberforce Road, Cambridge CB3 0WA, UK}
 \affiliation{Shanghai Astronomical Observatory, Chinese Academy of Sciences, Shanghai 200030, China}
\author{Gordon I. Ogilvie}
 \email{gio10@cam.ac.uk}
\affiliation{Department of Applied Mathematics and Theoretical Physics, University of Cambridge, Centre for Mathematical Sciences, Wilberforce Road, Cambridge CB3 0WA, UK}



\begin{abstract}
Astrophysical disks that are sufficiently cold and dense are linearly unstable to the formation of axisymmetric rings as a result of the disk's gravity. In practice, spiral structures are formed, which may in turn produce bound fragments. We study a nonlinear dynamical path that can explain the development of spirals in a local model of a gaseous disk on the subcritical side of the gravitational instability bifurcation. Axisymmetric equilibria can be radially periodic or localized, in the form of standing solitary waves. The solitary solutions have an energy slightly larger than a smooth disk. They are further unstable to non-axisymmetric perturbations with a wide range of azimuthal wavenumbers. The solitary waves may act as a pathway to spirals and fragmentation.
\end{abstract}



\section{Introduction} \label{sec:intro}

Spirality is ubiquitous in astrophysics. The grand design spirals in galaxies and some circumstellar disks are believed to be global spiral density waves. Spirals can further regulate the formation of stars in galaxies \citep{Roberts1969large} and collapse to form planets in circumstellar disks \citep[see, e.g.][]{Durisen2007, Deng2021formation}. In this Letter we study the development of spiral structures through the gravitational instability (GI) of a massive gaseous disk orbiting in a central potential.

The physics of GI can readily be appreciated in a local patch of a thin gaseous disk \citep{GL1965}. In this 2D local model, $x$ and $y$ correspond to the radial and azimuthal directions. Fluid orbiting the centre at angular velocity $\Omega(r) \, \bm{\hat{e}_z}$ is governed by the equation of motion
\begin{equation}
\frac{D\bm{u}}{Dt}+2\Omega\,\bm{\hat{e}_z}\times \bm{u}=-\frac{\bm{\nabla}P}{\Sigma}-\bm{\nabla}(\Phi+\Phi_\text{t}),
\end{equation}
where $\bm{u}$ is the velocity, $D/Dt$ is the material derivative, $\Sigma$ and $P$ are the vertically integrated density and pressure, $\Phi$ is the gravitational potential due to gas self-gravity, and $\Phi_\text{t}=-\Omega Sx^2$ is the tidal potential. For uniform $\Sigma$ and $P$, the steady solution $\bm{u}_0=-Sx\,\bm{\hat{e}}_y$ represents the basic orbital motion with shear rate $S=-d\Omega/d\ln r$, which equals $3\Omega/2$ for a Keplerian disk dominated by a central mass. Hereafter we consider only the 
deviation from this background shear flow, 
$\bm{v}=\bm{u}-\bm{u}_0$.
\begin{figure*}
\centering
\includegraphics[scale=0.65]{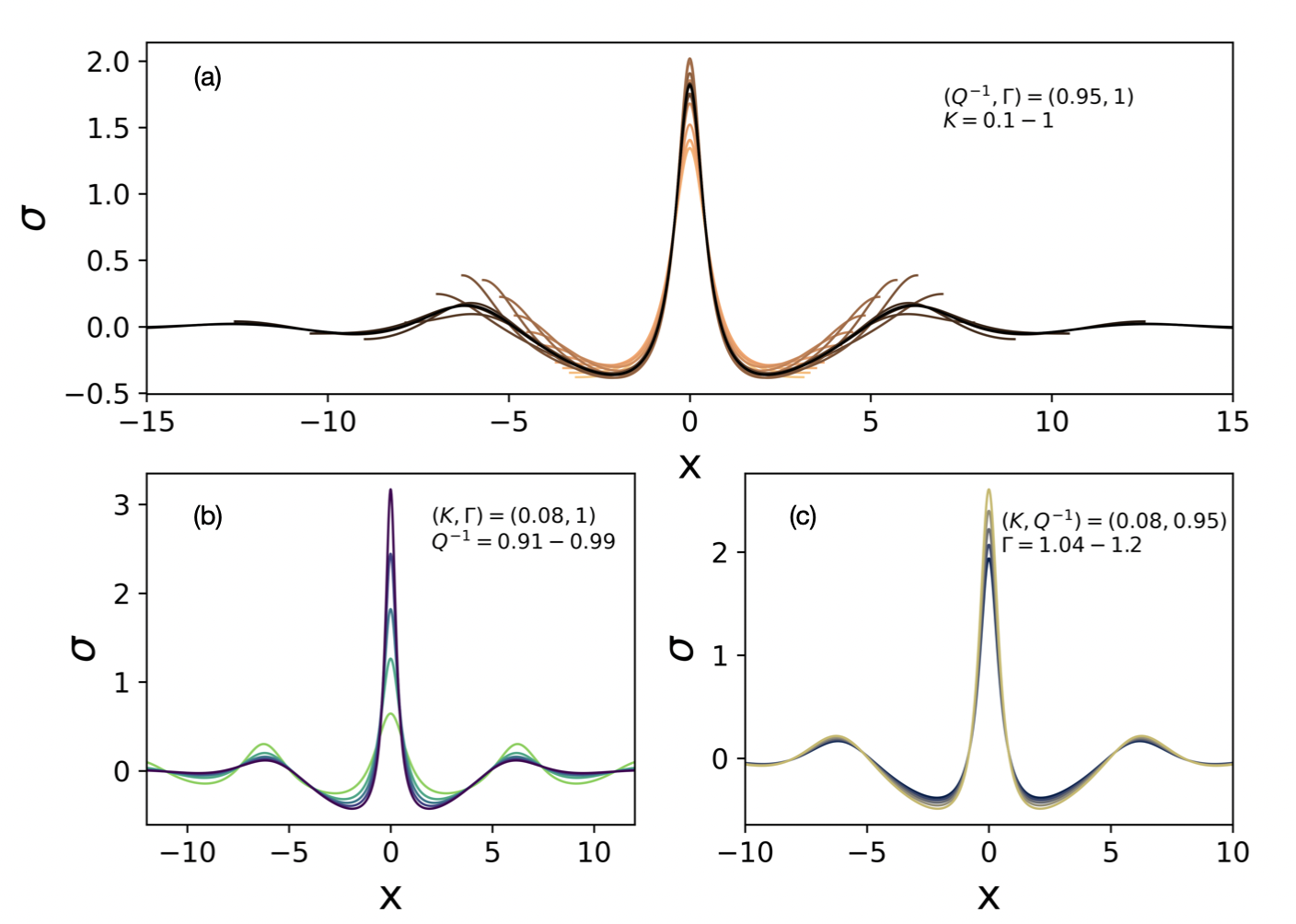}
\caption{Nonlinear periodic ring structures and solitary waves. (a) Periodic structures (only one wavelength is shown) with base wavenumber $K$ in an isothermal disk with $Q^{-1}=0.95$.  As $K$ decreases, the solutions transition to a solitary pattern (only the central part is shown here). (b,c) Solitary waves for different $Q^{-1}$ and different $\Gamma$ parameters, focusing on the central part of the solitary waves. The different line colors sweep the $K, Q^{-1}$, or $\Gamma$ parameters at constant rates, with the cooler colors representing smaller coefficients.}
\label{fig:soliton}
\end{figure*}

Axisymmetric perturbations with radial wavenumber $k_x$ oscillate at angular frequency $\omega$ given by
\begin{equation}
\omega^2=\kappa^2 - 2\pi G \Sigma|k_x| + c^2 k_x^2,
\end{equation}
where $c$ is the sound speed.
Long-wavelength (small $k_x$) perturbations are stabilized by the $\kappa^2=2\Omega(2\Omega-S)$ term (squared epicyclic frequency) resulting from conservation of angular momentum, while short-wavelength perturbations are stabilized by gas pressure \citep{[{e.g., review by ~}][]Shu2016}. A band of intermediate-scale axisymmetric perturbations exists when $Q=\kappa c /\pi G \Sigma<1$ \citep{Toomre1964}. The growth of non-axisymmetric spirals is less well understood and possibly linked to the overreflection of waves at the location that corotates with the spiral pattern.
If somehow the reflected wave is deflected again toward the corotation region, a feedback loop can be established so that the wave grows exponentially \citep{Mark1976, Nakagawa1992, Shu2016}. However, this mechanism only works efficiently for disks hovering on the brink of GI, and with enough radial structure to reflect the waves.
In 3D numerical simulations of gaseous disks, spirals start to grow at $Q\approx 1.5$ \citep{Durisen2007} and ring structures are often observed as transitions to spirals \citep{Mayer2008, Hirose2019, Deng2017}. 

Inspired by these numerical simulations, we investigated nonlinear steady axisymmetric structures in the local model as transitions to spirality. We unexpectedly discovered a class of standing solitary waves in a nonlinear integro-differential equation describing the radial force balance. Self-gravity of the gas introduces nonlocality through a Hilbert transform, resembling the Benjamin--Ono equation which also admits solitons \citep{Benjamin1967internal,Ono1975algebraic}. The solitary waves are found to transition to spiral structures via a secondary instability. We solve for nonlinear axisymmetric structures in section \ref{sec:structures}, examine their energy and stability in sections \ref{sec:energy} and \ref{sec:stability}, respectively, and draw conclusions in section \ref{sec:con}.

\section{Nonlinear axisymmetric structures} 
\label{sec:structures}
Two material invariants of the ideal fluid model are the specific entropy and the potential vorticity (PV) $(2\Omega-S+\partial_x v_y-\partial_y v_x)/\Sigma$. Throughout this paper we consider accessible solutions with the same uniform entropy, PV, and mean density as a uniform sheet (denoted by a subscript zero). For a perfect gas of (2D) adiabatic index $\Gamma>1$, isentropy implies the polytropic relations $P=A\Sigma^\Gamma$ and $dP/\Sigma=dW$, with specific enthalpy $W=A\Gamma\Sigma^{\Gamma-1}/(\Gamma-1)=c^2/(\Gamma-1)$. Hydrostatic 3D disks have an effective $\Gamma$ between $(3\gamma-1)/(\gamma+1)$ and $3-2/\gamma$ \citep{Gammie2001}, i.e., $1.33<\Gamma<1.57$ for a cold disk of molecular hydrogen with $\gamma=1.4$. 
Lower effective values of $\Gamma$ may be relevant when radiative processes are taken into account, with $\Gamma=1$ corresponding to the isothermal limit of instantaneous thermal relaxation.

For steady axisymmetric solutions ($v_x=0$), the PV constraint implies
\begin{equation}
\partial_x v_y=\frac{\kappa^2}{2\Omega}\sigma, \label{eqn:pv}
\end{equation}
where $\sigma=\Sigma/\Sigma_0-1$ is the fractional density variation. Radial force balance reads
\begin{equation}
-2\Omega v_y=-\partial_xW-\partial_x\Phi,
\end{equation}
and combines with (\ref{eqn:pv}) to give
\begin{equation}
\partial_x^2(W+\Phi)=\kappa^2 \sigma.
\label{eqn:rf}
\end{equation}
The dimensionless enthalpy perturbation
\begin{equation}
\frac{W-W_0}{c_0^2}=w=\frac{(1+\sigma)^{\Gamma-1}-1}{\Gamma-1} \label{eqn:enthalpy}
\end{equation}
reduces to $w=\text{ln}(1+\sigma)$ in the isothermal limit ($\Gamma \rightarrow 1$). $\Phi$ is related to $\Sigma$ through Poisson's equation in 3D, \textbf{regarding the disk as razor-thin}. This can be solved after taking a Fourier transform in $x$ and $y$, with the result
\begin{equation}
\tilde \Phi =-\frac{2\pi G}{k}\tilde{\Sigma}
\label{eqn:poisson}
\end{equation}
for any wavenumber $k=\sqrt{k_x^2+k_y^2}\neq  0$. In Fourier space, equation (\ref{eqn:rf}), when combined with (\ref{eqn:poisson}), gives
\begin{equation}
(2\pi G\Sigma_0 |k_x|-\kappa^2)\tilde{\sigma}=c_0^2k_x^2\tilde{w}.
\end{equation}
We adopt $c_0/\kappa$ as a natural unit of length. Radial force balance can be expressed in real space as
\begin{equation}
w_{xx}+2Q^{-1}\mathcal{H}\sigma_x=\sigma,
\end{equation}
where $Q^{-1}=\pi G \Sigma_0/\kappa c_0$ and $\mathcal{H}$ is the Hilbert transform \citep{King2009}. The polytropic relation (\ref{eqn:enthalpy}) makes the problem nonlinear. We look for periodic solutions with a basic wavenumber $K$ in the form of cosine series, 
\begin{equation}
\sigma=\sum_{n=1}^{\infty}\sigma_n \cos(nKx), \quad w=\sum_{n=0}^{\infty}w_n \cos(nKx),
\end{equation}
so that the coefficients are related by 
\begin{equation}
(2Q^{-1}nK-1)\sigma_n=(nK)^2w_n.
\label{eqn:coeff}
\end{equation}

Asymptotic analysis is possible when the disk is nearly marginally stable, i.e., $Q^{-1}$ and $K$ are close to 1 with small increments $\delta (Q^{-1})$ and $\delta K$. For weakly nonlinear solutions,
\begin{equation}
2\delta(Q^{-1})-(\delta K)^{2}\approx-\frac{(2-\Gamma)(5-3\Gamma)}{8}\sigma_1^{2},
\end{equation}in which the left-hand side measures supercriticality with respect to the GI: it is positive when $Q<1$ and $K$ is sufficiently close to $1$. For $\Gamma<5/3$, which is expected, solutions exist on the \textit{subcritical} side of the bifurcation with $\delta (Q^{-1})<0$, i.e., $Q>1$. Using these approximate solutions as initial guesses, we solve equations (\ref{eqn:enthalpy}) and (\ref{eqn:coeff}) by Newton--Raphson iteration for a range of parameters $Q^{-1}$, $K$, and $\Gamma$. We utilize Fast Fourier Transform (FFT) algorithms with up to 5000 points to achieve good convergence when $K$ is small.

\begin{figure}[ht]
\includegraphics[scale=0.7]{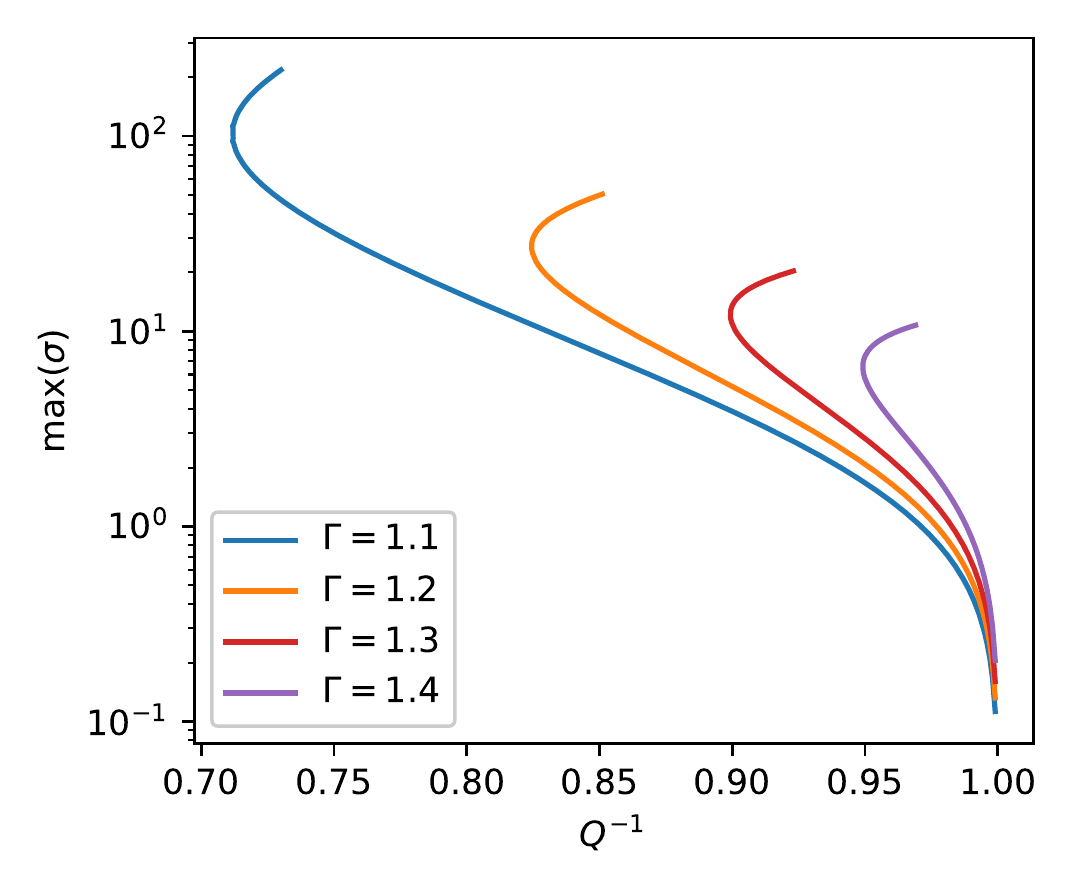}
\caption{The maxima of $\sigma$ as a function of $Q^{-1}$ when $K=1$.\label{fig:branch}}
\end{figure}

In Fig.~\ref{fig:soliton}, we plot the steady axisymmetric structures in various conditions. For a given $(Q^{-1},\Gamma)$ pair, the short-wavelength solutions ($K>1$) possess a single strong peak which tends to be infinitely sharp at large $K$, whereas the long-wavelength ($K<1$) solutions gradually approach a solitary wave. For example, in Fig.~\ref{fig:soliton}(a) the solitary wave is already established at $K=0.1$, and we observe no change in the pattern by tripling the base wavelength, $2\pi/K$. Hereafter the reported solitary waves for various ($Q^{-1},\Gamma$), whose shapes are well established, are all calculated with a base wavenumber $K=0.08$. A smaller $Q^{-1}$ or a larger $\Gamma$ leads to a more strongly peaked wave.

With a fixed wavenumber $K=1$ and $\Gamma<5/3$, we show the peak density perturbation of  periodic ring structures as a function of $Q^{-1}$ in Fig.~\ref{fig:branch}. The rings become more peaked when $Q^{-1}$ decreases from $1$ until the solutions turn onto an upper branch at a critical $Q^{-1}$ ($0.712$, $0.825$, $0.900$, $0.950$ for $\Gamma=1.1, 1.2,1.3,1.4$). The peak density continues to increase with $Q^{-1}$ until the first trough in the periodic structure (similar to Fig.~\ref{fig:soliton}) reaches zero and the upper branch ceases.

\section{Energy budget} \label{sec:energy}
\begin{figure*}
\includegraphics[scale=0.7]{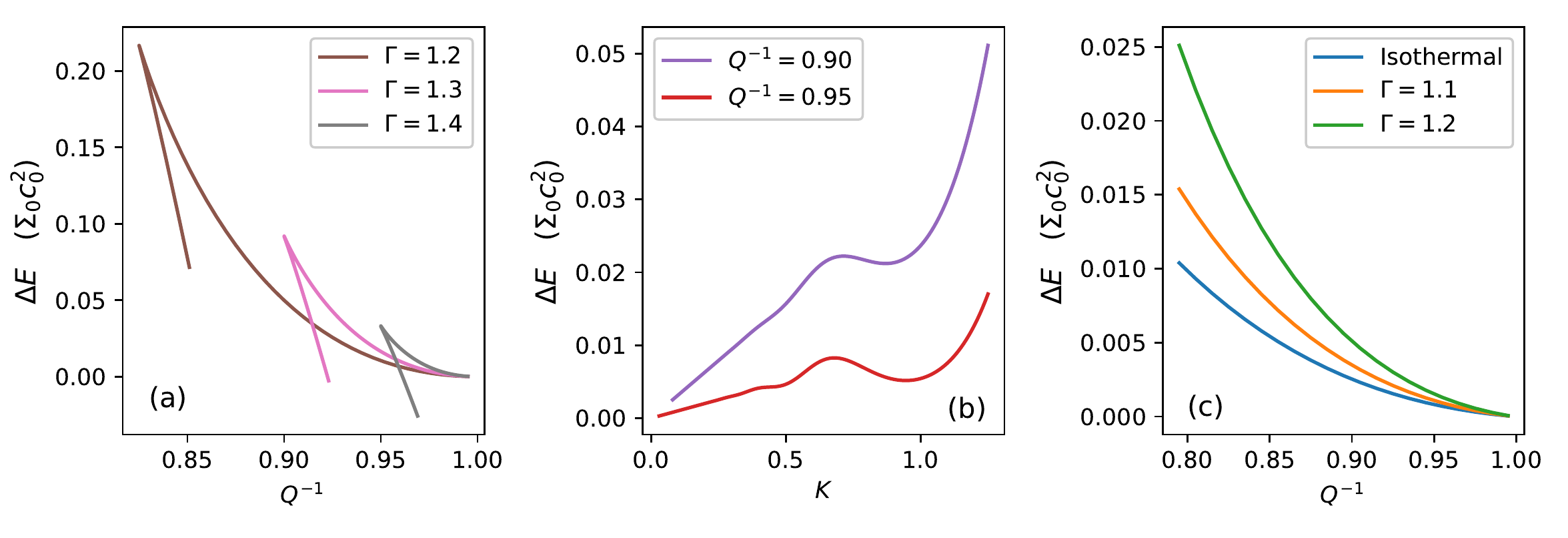}
\caption{\label{fig:energy} Energy change per unit area due to periodic rings or solitary waves. (a) Energy change for $K=1$ solutions. (b) Energy change as a function of the base wavenumber $K$ in isothermal disks. At sufficiently small $K$, $\Delta E/K$ tends to a constant because the solution transits to a solitary wave and the total energy change is invariant. (c) Energy change due to solitary waves (calculated with $K=0.08$). The total energy change per unit length in the $y$-direction is $78.54\, \Delta E$.}
\end{figure*}
The energy per unit area of the axisymmetric structures, relative to the uniform disk, is
\begin{equation}
\Sigma \left (\frac{1}{2}v_x^2+\frac{2\Omega^2}{\kappa^2}v_y^2\right)+\frac{1}{2}(\Sigma-\Sigma_0)\Phi+\frac{P-P_0}{\Gamma-1},
\end{equation}
where the first two terms combine the kinetic and tidal energies.
Considering the isovortical constraint (\ref{eqn:pv}) and radial force balance (\ref{eqn:rf}), and after integration by parts, assuming that $v_y$, $\sigma$, and $\Phi$ are either periodic or decaying in $x$, the total energy change due to the axisymmetric structure is 
\begin{equation}
\iint
\frac{\Sigma_0c_0^2}{\Gamma}\left [ w + \left(1-\frac{\Gamma}{2}\right)\sigma w\right] dx\,dy.
\end{equation}
So the mean energy change of periodic structures per unit area is 
\begin{equation}
\Delta E=\frac{\Sigma_0 c_0^{2}}{\Gamma}\left [w_0+\left(\frac{1}{2}-\frac{\Gamma}{4}\right)\sum_{n=1}^{\infty} \sigma_n w_n \right ].
\end{equation}

We plot the energy change due to periodic ring structures and solitary waves in Fig.~\ref{fig:energy}. For the $K=1$ solutions in Fig.~\ref{fig:branch}, the energy first increases as $Q^{-1}$ decreases along the lower branch as shown in Fig.~\ref{fig:energy}(a). This is expected for a subcritical bifurcation because the uniform disk is linearly stable and a local minimum of the energy. When the solution moves onto the upper branch, the energy turns over and falls with increasing $Q^{-1}$. The turnover points are stationary points of the energy \citep{Burke2006} and the upper branch solutions for $\Gamma>1.4$ can have energy lower than the undisturbed smooth sheet. The lower-branch  solutions are expected to be unstable.


In Fig.~\ref{fig:energy}(b), the energy change of the axisymmetric structure mostly increases with $K$ except around the local minima near $K=1$. In the long-wavelength limit ($K\rightarrow 0$), the energy perturbation per unit area scales linearly with $K$, i.e., $\Delta E/K$ tends to a constant. This is a sign that the solution converges to a solitary wave with a fixed total energy change. In Fig.~\ref{fig:energy}(c), the energy of solitary waves increases with $\Gamma$ and $Q$.

\begin{figure*}
\includegraphics[scale=0.45]{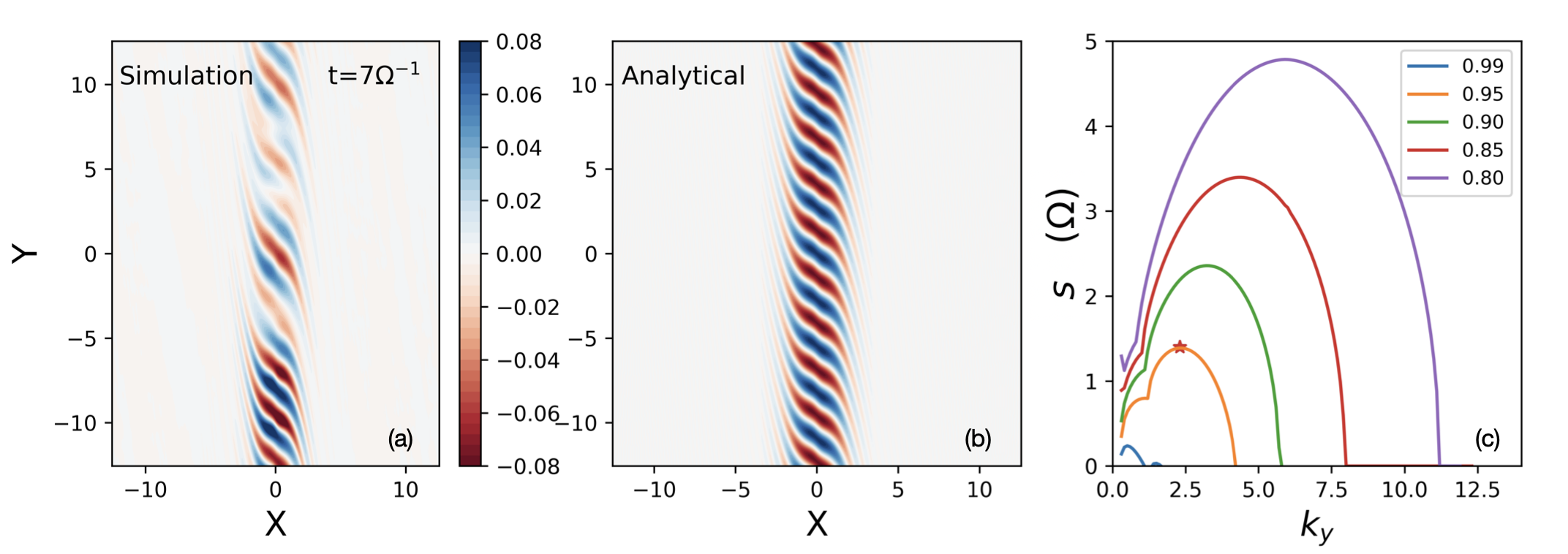}
\centering
\caption{\label{fig:modes} Non-axisymmetric modes attacking the solitary waves, and their growth rates. (a) Growing modes ($v^\prime_x$) attacking the solitary wave in a direct numerical simulation ($K=0.08$, only the central part plotted) of an isothermal disk with $Q^{-1}=0.95$. (b) Fastest-growing mode (Floquet analysis, arbitrary units) attacking the same solitary wave as in (a). (c) Growth rates of  modes with various $k_y$ in isothermal disks with solitary waves. The star symbol indicates the fastest growth rate measured in the numerical simulation in (a).}
\end{figure*}

\section{Non-axisymmetric instability}
\label{sec:stability}

We investigate the stability of the periodic and solitary waves to non-axisymmetric perturbations ($k_y\neq 0$). The evolution of linear perturbations (primed variables) is governed by
\begin{align}
\frac{Dv^\prime_x}{Dt} -2\Omega v^\prime_y&= -\partial_x(W^\prime+\Phi^\prime),
\\
\frac{Dv^\prime_y }{Dt} +(2\Omega-S+\partial_x v_y)v^\prime_x&=-\partial_y(W^\prime+\Phi^\prime),
\\
\frac{D\Sigma^\prime}{Dt}+v^\prime_x\partial_x\Sigma&=-\Sigma (\partial_x v^\prime_x+\partial_y v^\prime_y),
\end{align}
where $D/Dt = \partial_ t +(-Sx+v_y)\partial_y$ is the material derivative on the axisymmetric equilibrium. The perturbation can be expressed as a ladder of shearing waves \citep{GL1965}, e.g.,
\begin{equation}
\mathbf{v}^\prime=\sum_{n=-\infty}^{\infty}\mathbf{v}^\prime_n(t)\exp(ik_{x,n}x+ik_yy),
\end{equation}
with $k_{x,n}=nK+k_ySt$.  We only consider isovortical perturbations with zero potential vorticity change, i.e.,
\begin{equation}
\frac{\partial_xv^\prime_y-\partial_yv^\prime_x}{2\Omega-S}=\frac{\Sigma'}{\Sigma_0}.
\end{equation}

The density perturbation can thus be directly related to the velocity perturbations, leaving two coupled linear ordinary differential equations (ODEs) for each $n$:
\begin{widetext}
\begin{subequations}
\label{eqn:dvt}
\begin{equation}
\frac{dv^\prime_{x,n}}{dt}+ik_y\sum_{m=-\infty}^{\infty}v_{y,m}v^\prime_{x,n-m}-2\Omega v^\prime_{y,n}=-ik_{x,n}(W^\prime_n+\Phi^\prime_n),
\end{equation}\begin{equation}
\frac{dv^\prime_{y,n}}{dt}+ik_y\sum_{m=-\infty}^{\infty}v_{y,m}v^\prime_{y,n-m}+(2\Omega-S)v^{\prime}_{x,n}+\sum_{m=-\infty}^{\infty}(\partial_x v_y)_m v^{\prime}_{x,n-m}=-ik_y(W^\prime_n +\Phi^\prime_n).
\end{equation}\end{subequations}
\end{widetext}
Here, the Fourier coefficients of $v_y$ and $\partial_x v_y$ for the axisymmetric equilibrium can be easily obtained from the cosine series. The Fourier coefficients of $\Sigma^\prime$ can be related to the velocity perturbation coefficients by the isovortical condition. $\Phi^\prime_n$ is readily obtained from $\Sigma^\prime_n$ by the Poisson equation, and $W^\prime_n$ can be calculated via a convolution between $\Sigma^\prime_n$ and $(c^2/\Sigma)_n$.

The ODEs have $t$-dependent coefficients but the system is periodic in $t$, because the $k_{x,n}$ ladder shifts one place in $n$ after the recurrence time $T = K/k_yS$.  Hence Floquet's method can be used to determine the growth rates of various perturbations with $|n|$ not exceeding a truncation order $N$ \citep{Vanon2016}. To that end, a $2(2N+1)\times 2(2N+1)$ monodromy matrix was produced by applying $2(2N+1)$ sets of initial conditions where all variables but one are set to zero, with a different nonzero variable  in each case, to equation~(\ref{eqn:dvt}) and integrating it for one recurrence time.  To ensure the $\bm{v}^\prime_{\pm N}$ entries are properly evaluated, we extended the Fourier series, employing $2(N+N_0)+1$ wavenumbers for each variable in equation (\ref{eqn:dvt}). Here $N_0$ is the number of nonzero ($>10^{-14}$) modes in the equilibrium axisymmetric structures so that the convolutions in equation (\ref{eqn:dvt}) are properly handled for all modes $(-N\rightarrow N)$ in the monodromy matrix. We used up to $N=800$ wavenumbers for some strongly peaked solitary waves with $N_0$ up to 700. FFT and inverse FFT were utilized for the convolutions. 


The growth rate $s$ is determined by the eigenvalues $\lambda$ of the monodromy matrix (i.e., Floquet multipliers):
\begin{equation}
s=\frac{1}{T}\max{\mathrm{Re}(\ln\lambda)}.
\end{equation}
The corresponding eigenvectors give the velocity perturbations in the Fourier space. We note that there is typically only one growing mode for a given $k_y$. We focus on the stability of the solitary waves since they are localized and not affected by the radial boundary condition.
In Fig.~\ref{fig:modes}(c), we plot the growth rates of non-axisymmetric modes with different $k_y$ in isothermal disks. The sharper solitary waves at smaller $Q^{-1}$ are susceptible to more vigorous instabilities over a wider $k_y$ range. We note that a
larger value of $\Gamma$
narrows the window of unstable modes in $k_y$, but the fastest growth rate over all $k_y$ is only slightly affected.

The $v^\prime_x$ structure for the fastest growing mode with $Q^{-1}=0.95$ is shown in Fig.~\ref{fig:modes}(b). We verified our Floquet analysis by carrying out direct hydrodynamic simulations with the ATHENA code \citep{Stone2008}\footnote{The version 4.2 we downloaded from the official website has a wrong implementation of 2D self-gravity. It calculates gravity by the 3D gravitational stress tensor that does not apply to 2D systems. Instead, we modified the code and added the gravity force as a source term.} We initialized the solitary wave in a square sheet of size $2\pi/K$ and added random velocity noise of $0.1\%$ of the sound speed. The sheet is resolved by 8192 cells, i.e., 104 cells per $c_0/\kappa$, to minimize numerical dissipation, and we note that the instability grows slower in low-resolution simulations. In Fig.~\ref{fig:modes}(a), the simulation reproduces the velocity structure in the analytical calculation, and the measured fastest growing mode in the simulation has the predicted wavelength and growth rate. Fig.~\ref{fig:modes}(a) shows contamination by other modes close to $k_y\approx2.5$ because they have comparable growth rates.   

The solitary waves are expected to transition quickly to spiral structures. In our direct simulations of isothermal disks, the growing non-axisymmetric perturbation leads to runaway collapse of the solitary wave on scales of a few tenths of $c_0/\kappa$ \citep{Deng2021formation}. This is expected in isothermal environments and resolving the collapsing clumps becomes increasingly challenging. Unfortunately, solving the energy equation in the $\Gamma \neq 1$ case with our current hydrodynamical code \citep{Mullen2021} leads to inaccuracy, preventing us from directly simulating the expected  transition to spirals. Further analysis and simulations on the effects of vertical stratification and the equation of state are desirable.

\section{Conclusions.}
\label{sec:con}
 We investigated the structure of nonlinear axisymmetric equilibria on the subcritical side of the GI bifurcation. Periodic ring structures of a wide range of length scales exist. A class of solitary waves with slightly higher energy than the smooth disk is of particular interest, which may be induced by external perturbations. The solitary waves are unstable to non-axisymmetric perturbations, providing a new way of understanding the generation of spirals in subcritical self-gravitating accretion disks
\section{acknowledgments}
This research was funded by the Isaac Newton Trust (University of Cambridge) through research grant 21.07(d). HD also acknowledges support from the Swiss National Science Foundation via a postdoctoral fellowship. We thank the anonymous referee for suggestions that improved the clarity of the paper.

%

\vspace{5mm}


\software{SciPy \citep{Virtanen2020scipy},
ATHENA \citep{Stone2008}}
          





\bibliography{GIapj}{}
\bibliographystyle{aasjournal}



\end{document}